\documentclass[10pt,preprint2]{emulateapj}
\slugcomment{accepted to {\it Astrophysical Journal Letters}}
\shorttitle{Redshift evolution of the 1500\AA{} luminosity function}
\shortauthors{Arnouts et al.} 
\begin{document}
%
%
\title{The GALEX-VVDS\footnote{The VVDS observations have been
obtained with the European Southern Observatory Very Large
Telescope, Paranal Chile} \ \  Measurement of the Evolution of the 1500\AA{}
Luminosity Function}
%
\author{
S. Arnouts     \altaffilmark{1},
D. Schiminovich\altaffilmark{2,3}, 
O. Ilbert      \altaffilmark{1},      
L. Tresse      \altaffilmark{1},      
B. Milliard    \altaffilmark{1},
M. Treyer      \altaffilmark{1,3},   
S. Bardelli    \altaffilmark{17},
T. Budavari    \altaffilmark{4},    
T. K. Wyder    \altaffilmark{3},
E. Zucca       \altaffilmark{17},
O. Le F\`evre  \altaffilmark{1}, 
C. Martin      \altaffilmark{3}, 
G. Vettolani   \altaffilmark{10},
C. Adami       \altaffilmark{1}, 
M. Arnaboldi   \altaffilmark{15}, 
T. Barlow      \altaffilmark{3}, 
L. Bianchi     \altaffilmark{4},     
M. Bolzonella  \altaffilmark{18},
D. Bottini     \altaffilmark{11}, 
Y.-I. Byun     \altaffilmark{5}, 
A. Cappi       \altaffilmark{17}, 
S. Charlot     \altaffilmark{13,16}, 
T. Contini     \altaffilmark{12},
J. Donas       \altaffilmark{1},
K. Forster     \altaffilmark{3},     
S. Foucaud     \altaffilmark{11},  
P. Franzetti   \altaffilmark{11},
P. G. Friedman \altaffilmark{3}, 
B. Garilli     \altaffilmark{11}, 
I. Gavignaud   \altaffilmark{12},
L. Guzzo       \altaffilmark{14}, 
T. M. Heckman  \altaffilmark{4},  
C. Hoopes      \altaffilmark{1}, 
A. Iovino      \altaffilmark{14}, 
P. Jelinsky    \altaffilmark{6}, 
V. Le Brun     \altaffilmark{1},
Y.-W. Lee      \altaffilmark{5},      
D. Maccagni    \altaffilmark{11}, 
B. F. Madore   \altaffilmark{7}, 
R. Malina      \altaffilmark{1},     
B. Marano      \altaffilmark{18}, 
C. Marinoni    \altaffilmark{1},
H.J. McCracken \altaffilmark{16}, 
A. Mazure      \altaffilmark{1},
B. Meneux      \altaffilmark{1},
R. Merighi     \altaffilmark{17},
P. Morrissey   \altaffilmark{3},   
S. Neff        \altaffilmark{8}, 
S. Paltani     \altaffilmark{1},
R. Pell\`o     \altaffilmark{12}, 
J.P. Picat     \altaffilmark{12},  
A. Pollo       \altaffilmark{14}, 
L. Pozzetti    \altaffilmark{17},
M. Radovich    \altaffilmark{15}, 
R. M. Rich     \altaffilmark{9},    
R. Scaramella  \altaffilmark{10}, 
M. Scodeggio   \altaffilmark{11}, 
M. Seibert     \altaffilmark{3}, 
O. Siegmund    \altaffilmark{6}, 
T. Small       \altaffilmark{3},       
A. S. Szalay   \altaffilmark{4}, 
B. Welsh       \altaffilmark{6},
C. K. Xu       \altaffilmark{19}, 
G. Zamorani    \altaffilmark{17}, 
A. Zanichelli  \altaffilmark{10} 
       }
%
\email{stephane.arnouts@oamp.fr}
%
%
\altaffiltext{1}{Laboratoire d'Astrophysique de Marseille, BP 8, Traverse
du Siphon, 13376 Marseille Cedex 12, France}
\altaffiltext{2}{Department of Astronomy, Columbia University, MC2457,
550 W. 120 St. NY,NY 10027}
\altaffiltext{3}{California Institute of Technology, MC 405-47, 1200 East
California Boulevard, Pasadena, CA 91125}
\altaffiltext{4}{Department of Physics and Astronomy, The Johns Hopkins
University, Homewood Campus, Baltimore, MD 21218}
\altaffiltext{5}{Center for Space Astrophysics, Yonsei University, Seoul
120-749, Korea}
\altaffiltext{6}{Space Sciences Laboratory, University of California at
Berkeley, 601 Campbell Hall, Berkeley, CA 94720}
\altaffiltext{7}{Observatories of the Carnegie Institution of Washington,
813 Santa Barbara St., Pasadena, CA 91101}
\altaffiltext{8}{Laboratory for Astronomy and Solar Physics, NASA Goddard
Space Flight Center, Greenbelt, MD 20771}
\altaffiltext{9}{Department of Physics and Astronomy, University of
California, Los Angeles, CA 90095}
\altaffiltext{10}{INAF-IRA, Via Gobetti,101, I-40129, Bologna, Italy }
\altaffiltext{11}{INAF-IASF, via Bassini 15, I-20133, Milano, Italy }
\altaffiltext{12}{Laboratoire d'Astrophysique de l'Observatoire Midi-Pyr\'en\'ees,
 14, avenue E. Belin, F-31400 Toulouse, France }
\altaffiltext{13}{Max Planck Institut fur Astrophysik, 85741 Garching, Germany}
\altaffiltext{14}{INAF-Osservatorio Astronomico di Brera, via Brera 28,, Milan, Italy}
\altaffiltext{15}{INAF-Osservatorio Astronomico di Capodimonte, via Moiariello 16,
 I-80131 Napoli, Italy }
\altaffiltext{16}{Institut d'Astrophysique de Paris, UMR 7095, 98 bis bvd Arago,
 F-75014 Paris, France }
\altaffiltext{17}{INAF-Osservatorio Astronomico di Bologna, via Ranzani,1, I-40127 Bologna, Italy}
\altaffiltext{18}{Universit\`a di Bologna, Departimento di Astronomia,
 via Ranzani 1, I-40127 Bologna, Italy}
\altaffiltext{19}{NASA/IPAC, California Institute
of Technology, Mail Code 100-22, 770 S. Wilson Ave., Pasadena, CA 91125}
%
\newpage
\begin{abstract}
 We present the first measurement of the galaxy luminosity function at
 1500 \AA{} between $0.2 \le z \le 1.2$ based on GALEX-VVDS
 observations ($\sim$1000 spectroscopic redshifts for galaxies with
 $NUV \le 24.5$) and at higher $z$ using existing datasets.  Our main
 results are summarized as follows : 
 (i) luminosity evolution is observed with $\Delta M_{\star} \sim
 -2.0$ mag between $z=0$ and $z=1$ and $\Delta M_{\star}\sim -1.0$ mag
 between $z=1$ and $z=3$.  This confirms that the star formation
 activity was significantly higher in the past.
 (ii) the LF slopes vary between $-1.2 \ge \alpha \ge -1.65$, with a
 marginally significant hint of increase at higher $z$.
 (iii) we split the sample in three restframe $(B-I)$ intervals
  providing an approximate spectral type classification: Sb-Sd, Sd-Irr
  and unobscured starbursts. 
 We find that the bluest class evolves less strongly in luminosity than the
 two other classes. On the other hand their number density increases
 sharply with $z$ ($\sim$15\% in the local universe to
 $\sim$55\% at $z\sim 1$) while that of the reddest classes decreases.
\end{abstract}
%
%
\newpage
\section{Introduction}
%
 Evidence collected over the past decade suggests that the cosmic,
 volume-averaged star formation rate (SFR) density has increased from
 the present to $z\sim 1$ (Lilly et al. 1996, Wilson et al. 2002) with
 a possible flattening at higher redshift (Steidel et al. 1999,
 Giavalisco et al 2004).  The details of the evolution between $0\le
 z\le 1$ are still being debated and appear to depend on the selected
 wavelength.  Restframe far-ultraviolet (FUV: 1500\AA{}) observations
 can provide a sensitive measurement of the ongoing SFR but up to now
 have only been obtained in the local universe (2000\AA{}, Sullivan et
 al. 2000) and at $z\ge 2.5$ (1700\AA{}, Steidel et al.  1999). The
 GALEX mission now allows us to refine the local FUV measurements
 (Wyder et al. 2004, Treyer et al. 2004, Budavari et al. 2004) and to
 fill the redshift gap where most of the SFR evolution has taken
 place.

 In this paper we use a unique spectroscopic sample of $\sim$1000
 NUV-selected galaxies to derive the first measurement of the
 1500\AA{} LF between $0.2\le z\le 1.2$. We also use HDF data to
 extend our analysis to $z=3$.  These results allow us to address
 the evolution of the luminosity density and SFR, presented in a
 companion paper (Schiminovich et al. 2004).  Throughout the paper, we
 assume a flat $\Lambda$CDM cosmology with $\Omega_M=0.3$ and $H_0=70$
 km s$^{-1}$ Mpc$^{-1}$.
%
\section{Data description}\label{sec:obs}
%
   We use the data collected in the 02hr field
 (02h26m00s,$-04^{\circ}$30$^{'}$00$^{"}$) by GALEX and by the VIRMOS
 VLT Deep Survey (VVDS).
   The 02hr field was observed as part of the GALEX Deep Imaging
  Survey (DIS), with a total integration time of t$_{exp}=52763sec$ in
  two channels (FUV $\sim$1530\AA{} and NUV $\sim$2310\AA{}).  We
  restrict the analysis to the central $1^{\circ}$ diameter region of
  the GALEX field of view which provides higher uniformity (Morrissey
  et al. 2004).  The typical PSF has a $FWHM\sim 5$ arcsec. Because of
  the high source density in these images, the GALEX data pipeline
  (implementing a modified version of SExtractor, Bertin \& Arnouts
  1996) was run with deblending parameters configured to improve
  extraction of unresolved point sources at the faint limit.  The
  number counts are $\sim$80\% complete at 24.5 (AB system) in the FUV
  and NUV bands (Xu et al. 2004). Magnitudes were corrected for
  Galactic extinction using the Schlegel et
  al. (1998) reddening map and Cardelli et al. (1989) extinction
  law with $A_{FUV}/E(B-V)= 8.29$, $A_{NUV}/E(B-V)=8.61$.

 The VVDS is a spectroscopic survey with multicolor photometry
 (Le F\`evre et al 2003). Deep $B,V,R,I$ imaging over 1.2 deg$^2$ was
 carried out using the Canada-France-Hawaii telescope, with typical
 depths of 27,26,26.2,25.3 respectively and a PSF better than 1 arcsec
 (McCracken et al. 2003).  The ongoing deep spectroscopic survey
 covers $\sim$1950 arcmin$^2$ with a preliminary sample of $\sim$7200
  spectroscopic redshifts of galaxies with $17.5\le I_{AB} \le 24$.

  In the UV sample, all sources have at least one optical counterpart
 (hereafter OC) within a radius of 4$^{\prime\prime}$.  With the VVDS
 spectroscopic sample, we match 1157 galaxies with $NUV\le24.5$ and
 find that 48\% are isolated (one OC in 4$^{\prime\prime}$), 36\% have two OCs
 and 16\% have more than two OCs.  These fractions are almost constant
 with magnitude between $21.5\le NUV\le 24.5$ and with redshift.  As a
 preliminary matching procedure, we associate the closest optical
 counterpart to each UV source.  This also selects the brightest $B$
 band counterpart in 95\% of the double counterpart cases and 85\% of
 the more than two counterpart cases.  In the case of NUV sources
 with several OCs, the UV flux may result from the UV contribution of
 several sources.  Although we do not correct for the UV flux
 attributed to the chosen OC, we use simulations to estimate the
 magnitude of the bias.
 
  To test the impact of blends, we divide the NUV flux among the
  potential OCs determined using a match reliability
  statistic based on the likelihood estimator of Sutherland and
  Saunders (1992). We find that the average UV flux overestimate is
  $\sim -0.25$ mag in the case of two OCs and up to $\sim -0.5$ mag
  for the more complex cases. By comparing the luminosity functions
  derived for this blend-corrected NUV selected sample and the original
  sample, we find that the variations of $M_{\star}$ is less than
  $-0.25$, within the errorbars quoted below. This method may be more
  accurate but still requires several strong assumptions (optical
  and UV sources are intrinsically unresolved, optical-UV color is
  fixed). As we have shown that the results are not significantly
  affected by the adopted method, we use the simplest approach.

\begin{figure}
\includegraphics[angle=0,scale=.35]{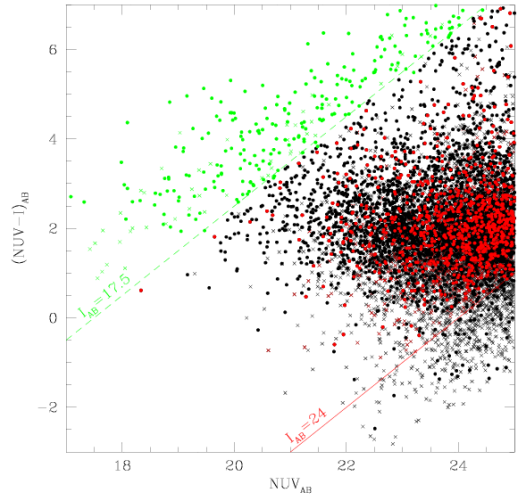}
\caption{ Color-magnitude diagram ($(NUV-I)_{AB}$ vs $NUV_{AB}$).
\label{fig:col}       }
\end{figure}

  Figure~\ref{fig:col} shows the $(NUV-I)$ vs $NUV$ color magnitude
 diagram. Isolated objects are displayed with filled circles and
 multiple counterpart cases with crosses.  The spectroscopic sources
 (red symbols) randomly sample the full range of colors observed in
 the total sample and is therefore unbiased.  For $I$ band saturated
 galaxies ($\sim$4\% with $I_{AB}\le17.5$, green symbols), we estimate
 that $\sim$95\% are at $z\le 0.2$ using photometric redshifts in
 GALEX-SDSS fields (Budavari et al. 2004). No redshifts were measured
 for galaxies with $I_{AB}\ge24$, selecting against very blue objects,
 but the small fraction ($\sim$4\%) is unlikely to affect our results.

 Although our preliminary catalog contains galaxies with spectroscopic
 redshifts between $0\le z \le 1.5$ we restrict the sample to the 1039
 galaxies in the redshift range between $0.2\le z\le 1.2$.  The low
 redshift cut avoids local galaxies saturated in the $I$ band.  The
 high redshift limit ensures that the the 912\AA{} break falls below
 the blue edge of the NUV passband.
%
\section {The FUV (1500\AA{}) luminosity functions} \label{sec:lf}
\begin{figure}
\includegraphics[angle=0,scale=.35]{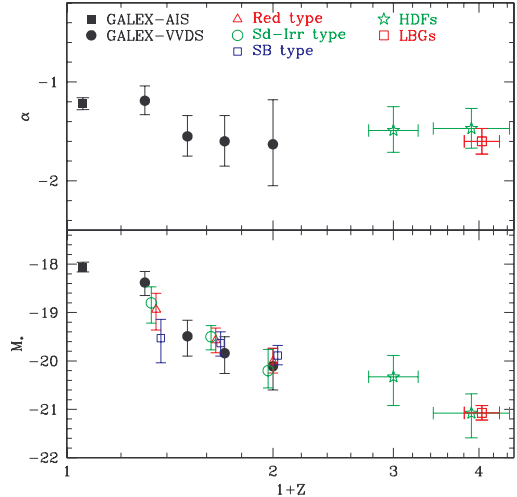}
\caption{Evolution of the LF's STY parameters vs redshift.
\label{fig:lfpar}  }
\end{figure}

%
 We derived restframe FUV absolute magnitudes using k-corrected NUV
 fluxes (with k-correction based on SED fits).  This choice yields
 minimal k-corrections because the average wavelength of the NUV
 passband samples the rest-FUV interval 1050\AA{}$\le \lambda_{rest}
 \le$ 1925\AA{} over our adopted redshift range.  To measure the
 luminosity function (LF) we use the VVDS LF tool (ALF) by Ilbert et
 al. 2004a, which includes the $1/V_{max}$, $C+$, SWML and STY
 estimators.  A weight is assigned to each spectroscopic galaxy to
 take into account the spectroscopic strategy of the VVDS observations
 and the completeness correction of the NUV number counts. We refer to
 a forthcoming paper for a complete description of the absolute
 magnitude and weight measurements.
%
\begin{figure}
\includegraphics[angle=0,scale=.45]{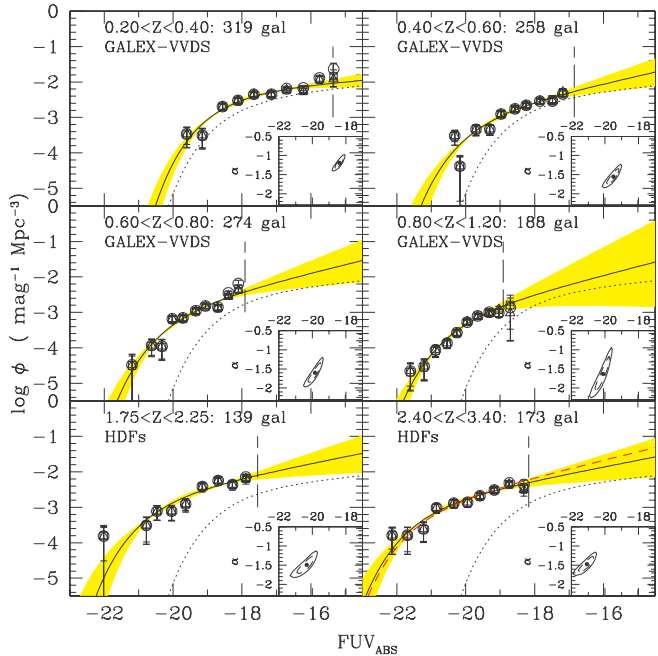}
\caption{1500\AA{} LF between $0.2\le z\le 3.5$.
\label{fig:lf}     }
\end{figure}
%
 Figure~\ref{fig:lf} shows the GALEX-VVDS LFs in four redshift bins (4
 upper panels) for the four estimators ($V_{max}$: circles, SWML:
 triangles, $C_+$: squares, STY: solid line). For the STY fits, we
 show the extrema (hashed area) based on the individual 1$\sigma$
 uncertainty in $\alpha$ and $M_{\star}$ (the likelihood probability
 contours are shown as insets for $2\Delta ln \mathcal L =1$: dashed
 lines and for $2\Delta ln \mathcal L =2.3$: solid lines).
  The vertical lines show the limits beyond which the global sample
 becomes biased against certain spectral types (see Ilbert et al.
 2004b). These values are listed in Table~\ref{tab:lfpar} as
 $M_{bias}$. Points beyond these limits are not included in the
 fitting procedures.  In general, NUV selection serves to reduce this
 bias significantly.  Finally we measure the FUV-LF at high-$z$ using
 photometric redshifts in the HDF North and South (Arnouts et al.
 2002; bottom panels).  For the purpose of restframe FUV selection, we
 define one sample between $1.75\le z\le 2.25$ using $F450$ passband
 with $F450_{AB}\le 27$ and one sample between $2.40\le z\le 3.40$
 using $F606$ passband with $F606_{AB}\le 27$.  At $z\sim3$, we also
 compare our results with the LF at 1700\AA{} (Steidel et al. 1999,
 red dashed line). As a reference, we show the local 1500\AA{} LF
 derived from GALEX data (Wyder at al. 2004; dotted lines).  All of
 the STY parameters are listed in Table~\ref{tab:lfpar} and
 Figure~\ref{fig:lfpar} shows the slope ($\alpha$, top panel) and the
 $M_{\star}$ (bottom panel) parameters versus redshift (GALEX-VVDS
 sample: filled circles, HDF sample: stars, LBG sample: open squares,
 local sample: filled squares).

 The faint end slopes vary between $-1.65 \le \alpha \le -1.2$ for
 $0.2\le z\le 3$ with a marginal steepening with $z$ (within the
 $1\sigma$ errorbars).  The $M_{\star}$ vs $z$ plot reveals strong
 redshift evolution.  A significant brightening of order $\Delta
 M_{\star}\sim -2$ occurs between $0\le z \le 1.2$.  The higher $z$
 samples show that the trend continues to $z\sim 3$ at a lower rate
 ($\Delta M_{\star}\sim-1$).  This increase is highly significant with
 respect to our errorbars and the source blending issue discussed in
 Section~\ref{sec:obs}.  The brightening of $M_{\star}$ and the
 steepening of the slope observed at 1500\AA{} is qualitatively
 consistent with the evolution detected at longer wavelengthes
 (2800\AA, U and B bands) by Wolf et al. (2003) and Ilbert et
 al. (2004a).

%
%

%
\begin{table}
\caption{ 1500\AA{} LF parameters for ($\Omega_0=0.3$,$\Omega_{\Lambda}=0.7$) and $H_0=70$ km s$^{-1}$
Mpc$^{-1}$
 \label{tab:lfpar}}
\begin{tabular}{cccccc}
\tableline
 z-bin        &Number&$M_{bias}$& $\alpha$        & $M_{\star}$              & $\Phi_{\star}^{a}$  \\
\tableline
 0.055$^{b}$   & 896 &  $--$    & $-1.21\pm 0.07$ & $-18.05\pm 0.11$         & $4.07 \pm 0.56$  \\
\tableline
0.2 - 0.4      & 319 & $-15.36$ & $-1.19\pm 0.15$ & $-18.38 \pm 0.25$        & $6.15 \pm 1.76$   \\
0.4 - 0.6      & 258 & $-16.85$ & $-1.55\pm 0.21$ & $-19.49 \pm 0.37$        & $1.69 \pm 0.88$   \\
0.6 - 0.8      & 274 & $-17.91$ & $-1.60\pm 0.26$ & $-19.84 \pm 0.40$        & $1.67 \pm 0.95$   \\
0.8 - 1.2      & 188 & $-18.92$ & $-1.63\pm 0.45$ & $-20.11 \pm 0.45$        & $1.14 \pm 0.76$   \\
\tableline
0.2 - 0.5$^{c}$& 137 & $--$     & $-1.40\pm 0.20$ & $-18.94^{+0.34}_{-0.42}$ & $0.99^{+0.58}_{-0.46}$ \\
0.5 - 0.8$^{c}$& 93  & $--$     & $-1.40\pm 0.20$ & $-19.57^{+0.22}_{-0.31}$ & $0.58^{+0.22}_{-0.20}$ \\
0.8 - 1.2$^{c}$& 59  & $--$     & $-1.40\pm 0.20$ & $-20.01^{+0.24}_{-0.27}$ & $0.45^{+0.10}_{-0.10}$ \\
\tableline
0.2 - 0.5$^{d}$& 153 & $--$     & $-1.50\pm 0.20$ & $-18.80^{+0.33}_{-0.42}$ & $0.90^{+0.56}_{-0.43}$ \\
0.5 - 0.8$^{d}$& 120 & $--$     & $-1.50\pm 0.20$ & $-19.50^{+0.23}_{-0.27}$ & $0.63^{+0.26}_{-0.23}$ \\
0.8 - 1.2$^{d}$& 28  & $--$     & $-1.50\pm 0.20$ & $-20.20^{+0.36}_{-0.44}$ & $0.16^{+0.04}_{-0.04}$ \\
\tableline
0.2 - 0.5$^{e}$& 152 & $--$     & $-1.50\pm 0.20$ & $-19.53^{+0.39}_{-0.51}$ & $0.60^{+0.45}_{-0.32}$ \\
0.5 - 0.8$^{e}$& 196 & $--$     & $-1.50\pm 0.20$ & $-19.63^{+0.23}_{-0.27}$ & $1.03^{+0.42}_{-0.37}$ \\
0.8 - 1.2$^{e}$& 101 & $--$     & $-1.50\pm 0.20$ & $-19.89^{+0.19}_{-0.21}$ & $0.80^{+0.18}_{-0.19}$ \\
\tableline
1.75-2.25$^{f}$& 139 & $-17.54$ & $-1.49\pm 0.24$ & $-20.33 \pm 0.50$        &  $2.65 \pm 2.00$  \\
2.40-3.40$^{f}$& 173 & $-18.17$ & $-1.47\pm 0.21$ & $-21.08 \pm 0.45$        &  $1.62 \pm 0.90$  \\
\tableline
2.50-3.50$^{g}$& 564 & $--$     & $-1.60\pm 0.13$ & $-21.07 \pm 0.15$        &  $1.40$           \\
\tableline
\end{tabular}
\tablenotetext{a}{: 10$^{-3}$ Mpc$^{-3}$ }
\tablenotetext{b}{: Local sample with $z\le0.1$ and $FUV\le20$ }
\tablenotetext{c}{: $(B-I)\ge 0.85$ sample with $NUV\le24.5$ }
\tablenotetext{d}{: $0.56\le (B-I) \le 0.85$   sample with $NUV\le24.5$ }
\tablenotetext{e}{: $(B-I)\le 0.56 $        sample with $NUV\le24.5$ }
\tablenotetext{f}{: HDF      sample with $F450\le27$ and $F606\le27$ }
\tablenotetext{g}{: LBG      sample with $R\le25$ }
\end{table}
%

\section{Dependence of the FUV LF on color} \label{sec:type}
 The luminosity function of galaxies has been shown to vary as a
 function of rest-frame color or spectral type (Blanton et al. 2001,
 Wolf et al. 2003).  In this section we explore the FUV LF and its
 evolution as a function of color.

\begin{figure}
\includegraphics[angle=0,scale=.35]{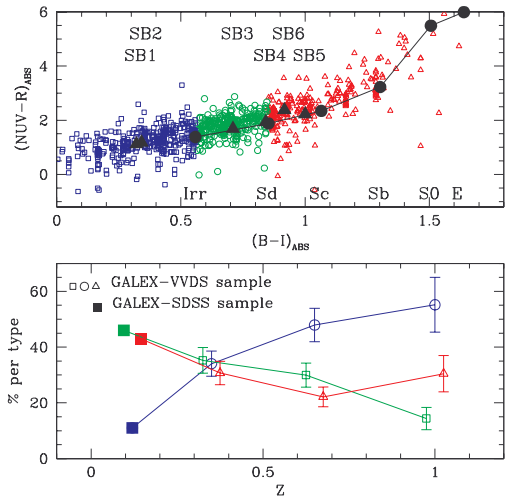}
\caption{ Restframe color-color plot for the color classification (upper panel).
Fraction of galaxy classes vs redshift (bottom panel).
\label{fig:type}  }
\end{figure}

 As shown by Salim et al. (2004), restframe $(NUV-R)$ is tightly
 correlated with the ratio of current to past averaged SFR, implying
 that the star formation history of a galaxy can already be
 constrained using a single color. In Fig.~\ref{fig:type}, we plot
 $(NUV-R)$ vs $(B-I)$ restframe as well as the location of theoretical
 SEDs from Elliptical to Sd (Poggianti 1997: filled circles), observed
 Irregular (Coleman et al. 1980: filled circle) and Starbursts (Kinney
 et al. 1996: filled triangles).

 As $(B-I)$ correlates with $(NUV-R)$ (Fig~\ref{fig:type}) and is not
 affected by the UV flux uncertainties mentioned above, we use $(B-I)$
 as a proxy for spectral types. We note however the degeneracy between
 dust and age of the stellar population:
  dusty starbursts (SB3 to SB6, filled triangles) cannot be
 distinguished from spiral galaxies (filled circles) using
 $(B-I)$. Eliminating this degeneracy will require additional color
 information sensitive to the old star population (e.g. $K$ band flux)
 not available in the present dataset. 

%
%

  We split our sample into three classes: $(B-I)\le 0.56$
 (corresponding to unobscured starbursts), $0.56\le (B-I) \le 0.85$
 (corresponding to Irr to Sd with a contamination of obscured
 starbursts) and $(B-I)\ge 0.85$ (corresponding to Sd to Sb with a
 contamination of obscured starbursts, and a negligible 7\%
 contribution of types earlier than Sb).  The shortage of red systems
 in our sample is consistent with the morphological analysis of de
 Mello et al. (2004) based on a small NUV restframe sample (34
 objects) in the CDF-South observed with HST (ACS). The authors only
 found 2 early type galaxies (8\%), the rest being late types (32\%)
 and starbursts (60\%).

 In the bottom panel of Fig.~\ref{fig:type}, we show the relative
 fraction for each color class in three redshift bins: $0.2\le z\le
 0.5$, $0.5\le z\le 0.8$, $0.8\le z\le 1.2$. We extend our analysis to
 lower $z$ by applying the same criteria to the GALEX-SDSS fields for
 galaxies with $z_{phot}\le 0.2$ and $FUV\le 22$ (filled squares).  We
 find that the contribution of unobscured starbursts rises from $\sim
 12\%$ at $z\sim 0.1$ to 55\% at $z\sim 1$, while the fractions of the
 two other classes decrease.  (The reddest class actually seems to
 increase again at $z\sim1$, but this may reflect a classification
 bias since the $I$ band probes the emitted light below the
 $4000$\AA{} break and does not trace reliably the old stellar
 population).

 We measure the LFs for the three classes. Due to the small number of
  galaxies we fixed the slopes to $\alpha=-1.4 \pm 0.2$ for our
  reddest class and $\alpha=-1.5 \pm 0.2$ for the two others, in
  agreement with the values obtained between $0.2\le z\le 0.5$.  These
  values are consistent with the slopes derived by Wolf et al. (2003)
  at 2800\AA{} for their types 2 to 4.  The best fit STY parameters
  are listed in Table~\ref{tab:lfpar} and in Fig~\ref{fig:lfpar}.

 We find that the bluest class (unobscured starbursts) evolves less
 strongly in luminosity than the two other classes ($\Delta M_{\star}
 \le -0.5$ mag between $0.35\le z \le 1.0$ as opposed to $\Delta
 M_{\star} \ge -1$ mag).  On the other hand the number densities
 ($\rho=\int_{-\infty}^{M_{cut}} \Phi(M)dM$ with $M_{cut}=-18.5$)
 evolve similarly to the non-volume corrected fractions shown in
 Fig~\ref{fig:type}, namely the density of unobscured starbursts
 increases sharply with $z$ while that of the reddest classes
 decreases.

 The interpretation of our results in terms of star formation rate
 evolution is presented in a companion paper by Schiminovich et
 al. (2004).
%
\acknowledgments GALEX (Galaxy Evolution Explorer) is a NASA Small
 Explorer, launched in April 2003.We gratefully acknowledge NASA's
 support for construction, operation and science analysis for the GALEX
 mission, developed in cooperation with the Centre National d'Etudes
 Spatiales of France and the Korean Ministry of Science and
 Technology.\\
 The VVDS is supported by the Centre National de la Recherche
 Scientifique of France and its Cosmology program, the Observatoire
 Astronomique Marseille Provence, and by the Italian National Research
 Council.
%

%
%
%
%
\end{document}